%% file: liege08_demink.tex
\documentclass[mypaper,7pt,twoside]{CoAst}
\usepackage{epsf,graphicx,fancyhdr}
\usepackage{natbib}
\input{CoAst_liege-proceedingslayo}

\newcommand{\Msun}{\mbox{$~\mathrm{M}_{\odot}$}}

\def\apjl{\mbox{{ApJL}}}
\def\araa{\mbox{{ARAA}}}

\begin{document}
\sf

\chapterCoAst{ Rotational mixing in tidally locked massive main
sequence binaries }
{S.E.\,de\,Mink et al.} 
\Authors{S.E.\,de\,Mink, M.\,Cantiello, N.\,Langer, O.R.\,Pols}
\Address{ $^1$ Astronomical Institute Utrecht, Princetonplein 5, 3584
CC Utrecht, The Netherlands, S.E.deMink@uu.nl \\ }

\noindent
\begin{abstract}
One of the main uncertainties in evolutionary calculations of massive
stars is the efficiency of internal mixing. It changes the chemical
profile inside the star and can therefore affect the structure and
further evolution. 

We demonstrate that eclipsing binaries, in which the tides synchronize
the rotation period of the stars and the orbital period, constitute a
potentially strong test for the efficiency of rotational mixing. We
present detailed stellar evolutionary models of massive binaries
assuming the composition of the Small Magellanic Cloud. In these
models we find enhancements in the surface nitrogen abundance of up to
0.6 dex.
\end{abstract}

\section{Introduction}
The inclusion of rotation into stellar evolution models has been shown
to be very successful in explaining various observed characteristics
of stars \citep[see][for a review]{Maeder+Meynet+review00}. It can
have a large effect on the internal distribution of elements, as it
leads to instabilities in the star, resulting in internal mixing.

The two most important mixing processes induced by rotation are
\emph{Eddington-Sweet circulations}, which consist of large scale
meridional currents originating from a thermal imbalance between pole
and equator in rotating stars \citep{Vonzeipel24, Eddington25,
Eddington26, Vogt25} and \emph{shear mixing}, which results from
eddies formed between two layers of the star rotating at different
angular velocities.

Near the center of a massive main sequence star hydrogen is converted
into helium, and carbon and oxygen into nitrogen.  Rotational mixing
can bring this processed material to the surface, where it can be
observed in the stellar spectra. Therefore rotation has been proposed
as an explanation for the enhanced nitrogen abundances observed in a
fraction of massive early type stars
\citep[e.g.][]{Walborn76,Maeder+Meynet+review00,Heger+Langer00}.

Although the effects of rotation on stellar evolution have been
 studied by various authors, we are still left with many questions. An
 essential question concerns the efficiency of rotational mixing.
 Attempts to constrain it have remained inconclusive due to limited
 sample sizes and/or a strong bias towards stars with small projected
 rotational velocities \citep[e.g.][]{Gies+Lambert92, Fliegner+96,
 Daflon+01, Venn+02, Korn+02, Huang+Gies06,Mendell+06}.

The recent VLT-flames survey of massive stars \citep{Evans+05}
provided for the first time a large sample of massive stars covering a
wide range of projected rotational velocities with accurate abundance
determinations \citep{Hunter+08rotandN}. Brott et al (2008, this Vol.)
demonstrated that the properties of the VLT-flames sample cannot be
reproduced by simulations of a population of rotating single stars.
This raises the question whether other processes, beside rotational
mixing, play an important role in explaining helium and nitrogen
enhancements of massive main sequence stars.  For example, the
observed enhancements could also be explained by binary interactions
\citep{Langer+proceedingchina08}. In this case a downward revision of
efficiency of the rotational mixing in single stars might be required.

Clearly, a strong and conclusive observational test for the efficiency
of rotational mixing is needed.  In this contribution we propose to
use detached eclipsing binaries for this purpose.

\section{Eclipsing binaries as laboratories for rotational mixing}

Eclipsing binaries have frequently been used to test stellar evolution
models as they provided the only method (until the development of
asteroseismological techniques) for accurate determinations of stellar
masses masses, radii and effective temperatures.  Even beyond our own
Galaxy, in the Magellanic Clouds, masses of O and early B stars have
been determined with accuracies of 10\% \citep{Hilditch+05}.
Rotational mixing is more important in more massive stars
\citep[e.g.][]{Heger+00}. Therefore, it is very useful to know the
stellar mass for quantitative testing of the efficiency of rotational
mixing: it enables a direct comparison with a corresponding stellar
evolution model with the proper mass.

In close binaries, with orbital periods ($P_{\rm orbit}$) less than a
few days, the tides are so strong that the stars rotate synchronously
with the orbital period: $P_{\rm spin} = P_{\rm orbit}$. With the
stellar radii known from eclipse measurements, this enables us to
determine the rotation rate directly from the orbital period. This is
an important advantage of using binaries for testing rotational mixing
with respect to single stars, for which fitting of spectral lines
allows only for the determination of $v \sin i$, where $v$ is the
rotational velocity at the equator.  The inclination $i$ of the
rotation axis is generally not known.

Here, we propose to use detached eclipsing binaries consisting of two
main sequence stars residing within their Roche lobes. Detailed
calculations of binary evolution show that if one of the stars fills
its Roche lobe during the main sequence, it does not detach again
before hydrogen is exhausted in the core, except maybe for a very
short thermal timescale \citep{wellstein+01, demink+07}. If we turn
the argument around we find that, in a binary with two detached main
sequence stars, we can safely exclude the occurrence of mass transfer
since the onset of core H burning.  The stars have lived their lives
similar to rotating single stars.  This is a third major advantage of
using eclipsing binaries with respect to single stars.  A fast
rotating single star may in contrast be the result of a merger of two
stars in a former binary. Moreover an apparently single star may have
been affected by mass transfer, while its companion may be very hard
to detect, being a faint low mass star in a wide orbit.

To test rotational mixing we need determinations of the surface
abundances.  If the spectra of a binary are of high quality, one can
determine the surface abundances, as is done for single stars, after
disentangling the composite spectra \citep[e.g.][]{Leushin88N,
Pavlovski+Hensberge05, Rauw+05}.  In the remainder of this paper we
discuss what type of binaries are suitable for testing rotational
mixing.

\section*{Stellar evolution code}

To investigate to what extent the surface abundances of close detached
binaries are affected by rotational mixing we model their evolution
using a detailed 1D stellar evolution code, described by
\citet{Yoon+06}, which includes the effects of rotation on the stellar
structure, the transport of angular momentum and chemical species via
rotationally induced hydrodynamic instabilities \citep{Heger+00} and
magnetic torques \citep{Spruit02, Heger+05}. Brott et al. (2008, this Vol.)
calibrated the efficiency of mixing processes in single star models
using the data from the VLT-flames survey \citep{Hunter+08rotandN}.

Tidal interaction is implemented as described in \citet{Detmers+08}
using the timescale for synchronization given by
\citet[eq. 6.1]{Zahn77}. The tides act on the outer layers of the
star. Angular momentum is redistributed in the stellar interior by
magnetic coupling and rotational instabilities. We note that the
binary systems modeled here are so tight that the stars are
synchronized throughout their main sequence evolution.

\figureCoAst{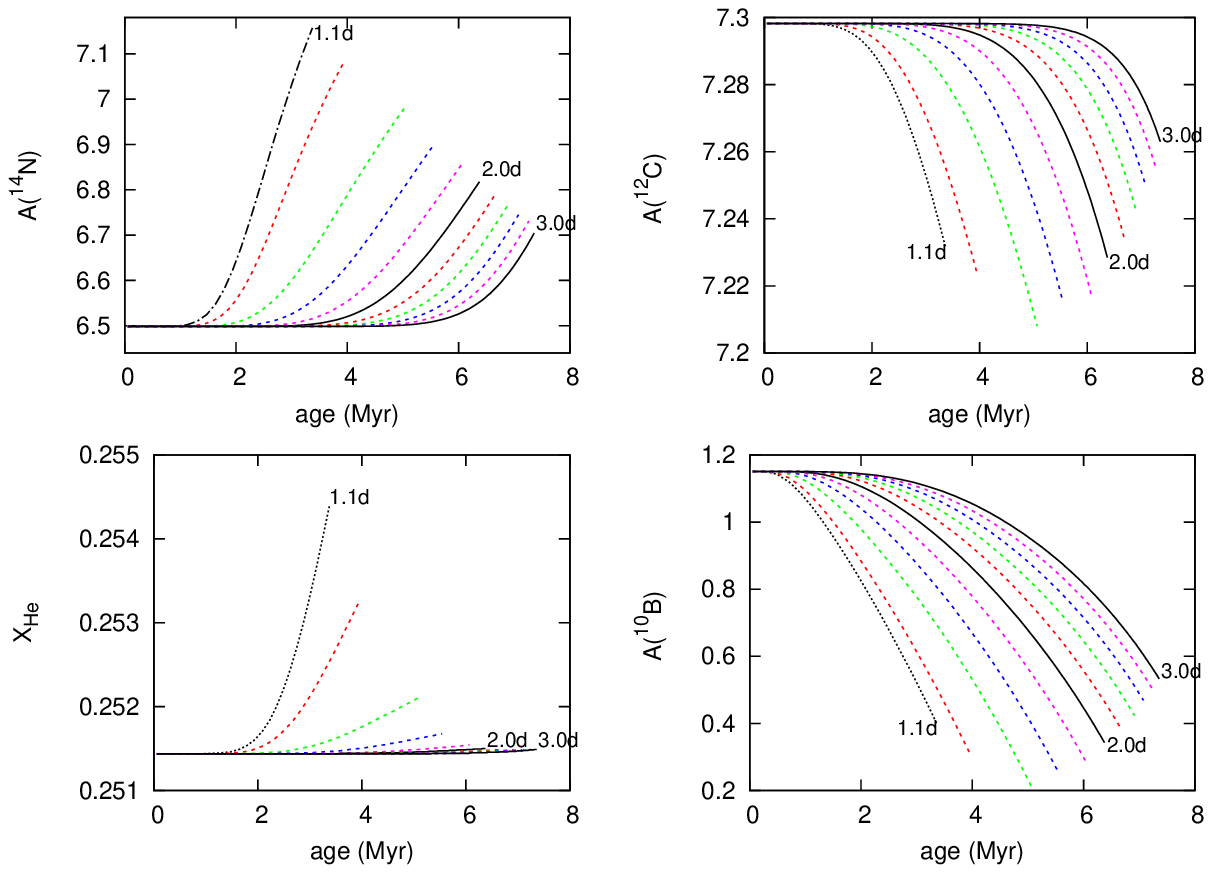}{ Surface abundances of nitrogen ($^{14}N$),
  carbon ($^{12}C$), boron ($^{10}B$) and the mass fraction of helium
  at the surface versus time for a 20\Msun~star with a 15\Msun~close
  companion. Note the different vertical scales. The abundance of an
  element X is given in the conventional units: $A({\rm X}) =
  \log_{10} (n_{\rm X} / n_{\rm H}) + 12$, where $n_X$ and $n_H$ refer
  to the number fractions.  The different lines show the evolution
  assuming different initial orbital period, varying between 1.1 and 3
  days.  The tracks are plotted from the onset of central H burning
  until the onset of Roche lobe
  overflow.}{surfabund}{htb}{clip,angle=0,width=\textwidth}

\section*{Results}

With the Small Magellanic Cloud (SMC) sample of  \citet{Hilditch+05} in
mind, which contains 21 detached systems%
     \footnote{Possibly only 20 systems are detached.  For two of the
       systems an alternative semi-detached solution.  For one of
       these systems a comparison to binary evolution models including
       the effects of mass transfer showed that the semi-detached
       solution was more consistent than the detached solution
       \citep{demink+07}.}
 with orbital periods of a few days and masses of the primary
 component up to 20\Msun, we chose to model the following binary
 systems. For the mass of the primary component we adopt 20\Msun, for
 the secondary component 15\Msun~ and we adopt initial orbital periods
 of up to three days.  We assume a composition representative of the
 small Magellanic cloud, which is relatively metal-poor and has a high
 carbon to nitrogen ratio.  The evolution is followed from the onset
 of central hydrogen burning at zero age until the primary star fills
 its Roche lobe.

 In all computed models the tides are efficient enough to keep both
 stars in synchronous rotation with the orbit. The shorter the orbital
 period, the faster the rotation of the stars, the more efficient
 rotational mixing, the faster the surface abundances change with
 time. On the other hand, the systems with short orbital periods are
 tighter and therefore the stars will fill their Roche lobe at an
 earlier stage, leaving less time to modify their surface abundances.
 These effects are illustrated in Fig.~\ref{surfabund}.

 Nitrogen is produced in the core and in the layers just above, as
 carbon and oxygen are consumed. Due to rotational mixing the nitrogen
 surface abundance increases and the carbon abundance decreases
 accordingly (see Fig.~\ref{surfabund}). Helium is produced deeper
 inside the star on a much longer timescale (the nuclear timescale).
 Some helium can be mixed up, but the helium surface enhancements
 achieved in our models are very small (less than 1\%). Another
 element that acts as a tracer of rotational mixing is boron.  This
 element can only survive in the coolest outermost layers of the
 star. Rotational mixing will bring it to hotter layers where it is
 destroyed. It is one of the elements most sensitive to rotational
 mixing. It is, however, hard to observe due to its low abundance,
 especially in the metal poor SMC. The increase in the nitrogen
 abundance in our models is up to three times larger than the typical
 error bar for surface abundance measurements in the VLT-flames survey
 (0.2 dex).  Nitrogen may therefore be the most suitable element to
 test rotational mixing.

\section*{Conclusion }
We have argued that eclipsing binaries can provide a potentially
stringent test for the efficiency of rotational mixing in massive
stars.  The stellar parameters and rotation rate can be accurately
determined enabling direct comparison to stellar evolution
models. Therefore even one well determined system could be used as a
test case.  By performing detailed evolutionary calculations of close
massive binaries, we show that (with currently assumed rotational
mixing efficiencies) we expect nitrogen enhancements of up to 0.6 dex for
binaries such as those in the sample of \citet{Hilditch+05}.

At present, it is not clear whether the presence a of binary companion
can lead to extra mixing, on top of the rotationally induced
mixing. If so, our proposed test would still constrain the efficiency
of rotational mixing in single stars by providing an upper limit to
this quantity. This will be discussed in a forthcoming paper.


\bibliographystyle{aa}

\References{

\bibitem[{{Daflon} {et~al.}(2001){Daflon}, {Cunha}, {Butler}, \&
  {Smith}}]{Daflon+01}
{Daflon}, S., {Cunha}, K., {Butler}, K., \& {Smith}, V.~V. 2001, \apj, 563, 325

\bibitem[{{De Mink} {et~al.}(2007){De Mink}, {Pols}, \& {Hilditch}}]{demink+07}
{De Mink}, S.~E., {Pols}, O.~R., \& {Hilditch}, R.~W. 2007, \aap, 467, 1181

\bibitem[{{Detmers} {et~al.}(2008){Detmers}, {Langer}, {Podsiadlowski}, \&
  {Izzard}}]{Detmers+08}
{Detmers}, R.~G., {Langer}, N., {Podsiadlowski}, P., \& {Izzard}, R.~G. 2008,
  \aap, 484, 831

\bibitem[{{Eddington}(1925)}]{Eddington25}
{Eddington}, A.~S. 1925, The Observatory, 48, 73

\bibitem[{{Eddington}(1926)}]{Eddington26}
{Eddington}, A.~S. 1926, {The Internal Constitution of the Stars} (Cambridge:
  Cambridge University Press, 1926)

\bibitem[{{Evans} {et~al.}(2005){Evans}, {Smartt}, {Lee}, {Lennon}, {Kaufer},
  {Dufton}, {Trundle}, {Herrero}, {Sim{\'o}n-D{\'{\i}}az}, {de Koter},
  {Hamann}, {Hendry}, {Hunter}, {Irwin}, {Korn}, {Kudritzki}, {Langer},
  {Mokiem}, {Najarro}, {Pauldrach}, {Przybilla}, {Puls}, {Ryans}, {Urbaneja},
  {Venn}, \& {Villamariz}}]{Evans+05}
{Evans}, C.~J., {Smartt}, S.~J., {Lee}, J.-K., {et~al.} 2005, \aap, 437, 467

\bibitem[{{Fliegner} {et~al.}(1996){Fliegner}, {Langer}, \&
  {Venn}}]{Fliegner+96}
{Fliegner}, J., {Langer}, N., \& {Venn}, K.~A. 1996, \aap, 308, L13

\bibitem[{{Gies} \& {Lambert}(1992)}]{Gies+Lambert92}
{Gies}, D.~R. \& {Lambert}, D.~L. 1992, \apj, 387, 673

\bibitem[{{Heger} \& {Langer}(2000)}]{Heger+Langer00}
{Heger}, A. \& {Langer}, N. 2000, \apj, 544, 1016

\bibitem[{{Heger} {et~al.}(2000){Heger}, {Langer}, \& {Woosley}}]{Heger+00}
{Heger}, A., {Langer}, N., \& {Woosley}, S.~E. 2000, \apj, 528, 368

\bibitem[{{Heger} {et~al.}(2005){Heger}, {Woosley}, \& {Spruit}}]{Heger+05}
{Heger}, A., {Woosley}, S.~E., \& {Spruit}, H.~C. 2005, \apj, 626, 350

\bibitem[{{Hilditch} {et~al.}(2005){Hilditch}, {Howarth}, \&
  {Harries}}]{Hilditch+05}
{Hilditch}, R.~W., {Howarth}, I.~D., \& {Harries}, T.~J. 2005, \mnras, 357, 304

\bibitem[{{Huang} \& {Gies}(2006)}]{Huang+Gies06}
{Huang}, W. \& {Gies}, D.~R. 2006, \apj, 648, 591

\bibitem[{{Hunter} {et~al.}(2008){Hunter}, {Brott}, {Lennon}, {Langer},
  {Dufton}, {Trundle}, {Smartt}, {de Koter}, {Evans}, \&
  {Ryans}}]{Hunter+08rotandN}
{Hunter}, I., {Brott}, I., {Lennon}, D.~J., {et~al.} 2008, \apjl, 676, L29

\bibitem[{{Korn} {et~al.}(2002){Korn}, {Keller}, {Kaufer}, {Langer},
  {Przybilla}, {Stahl}, \& {Wolf}}]{Korn+02}
{Korn}, A.~J., {Keller}, S.~C., {Kaufer}, A., {et~al.} 2002, \aap, 385, 143

\bibitem[{{Langer} {et~al.}(2008){Langer}, {Cantiello}, {Yoon}, {Hunter},
  {Brott}, {Lennon}, {de Mink}, \& {Verheijdt}}]{Langer+proceedingchina08}
{Langer}, N., {Cantiello}, M., {Yoon}, S.-C., {et~al.} 2008, in IAU Symposium,
  Vol. 250, IAU Symposium, 167--178

\bibitem[{{Leushin}(1988)}]{Leushin88N}
{Leushin}, V.~V. 1988, Soviet Astronomy, 32, 430

\bibitem[{{Maeder} \& {Meynet}(2000)}]{Maeder+Meynet+review00}
{Maeder}, A. \& {Meynet}, G. 2000, \araa, 38, 143

\bibitem[{{Mendel} {et~al.}(2006){Mendel}, {Venn}, {Proffitt}, {Brooks}, \&
  {Lambert}}]{Mendell+06}
{Mendel}, J.~T., {Venn}, K.~A., {Proffitt}, C.~R., {Brooks}, A.~M., \&
  {Lambert}, D.~L. 2006, \apj, 640, 1039

\bibitem[{{Pavlovski} \& {Hensberge}(2005)}]{Pavlovski+Hensberge05}
{Pavlovski}, K. \& {Hensberge}, H. 2005, \aap, 439, 309

\bibitem[{{Rauw} {et~al.}(2005){Rauw}, {Crowther}, {De Becker}, {Gosset},
  {Naz{\'e}}, {Sana}, {van der Hucht}, {Vreux}, \& {Williams}}]{Rauw+05}
{Rauw}, G., {Crowther}, P.~A., {De Becker}, M., {et~al.} 2005, \aap, 432, 985

\bibitem[{{Spruit}(2002)}]{Spruit02}
{Spruit}, H.~C. 2002, \aap, 381, 923

\bibitem[{{Venn} {et~al.}(2002){Venn}, {Brooks}, {Lambert}, {Lemke}, {Langer},
  {Lennon}, \& {Keenan}}]{Venn+02}
{Venn}, K.~A., {Brooks}, A.~M., {Lambert}, D.~L., {et~al.} 2002, \apj, 565, 571

\bibitem[{{Vogt}(1925)}]{Vogt25}
{Vogt}, H. 1925, Astronomische Nachrichten, 223, 229

\bibitem[{{von Zeipel}(1924)}]{Vonzeipel24}
{von Zeipel}, H. 1924, \mnras, 84, 665

\bibitem[{{Walborn}(1976)}]{Walborn76}
{Walborn}, N.~R. 1976, \apj, 205, 419

\bibitem[{{Wellstein} {et~al.}(2001){Wellstein}, {Langer}, \&
  {Braun}}]{wellstein+01}
{Wellstein}, S., {Langer}, N., \& {Braun}, H. 2001, \aap, 369, 939

\bibitem[{{Yoon} {et~al.}(2006){Yoon}, {Langer}, \& {Norman}}]{Yoon+06}
{Yoon}, S.-C., {Langer}, N., \& {Norman}, C. 2006, \aap, 460, 199

\bibitem[{{Zahn}(1977)}]{Zahn77}
{Zahn}, J.-P. 1977, \aap, 57, 383

}

\end{document}

%% file: CoAst_liege-proceedingslayo.tex
\renewcommand{\chaptermark}[1]{\markboth{#1}{}}

\newcommand{\apj}{ApJ}
\newcommand{\aap}{A\&A}  
\newcommand{\mnras}{MNRAS}

\newcommand{\kopf}{\small\itshape Comm. in Asteroseismology \\ Contribution to the Proceedings of the 38$^{th}$\,LIAC\,/\,HELAS-ESTA\,/\,BAG, 2008
}

\newcommand{\Authors}[1]{\begin{center}\normalsize\bf\sf #1 \end{center}}

\renewcommand{\author}[1]{\begin{center}\normalsize\bf\sf #1 \end{center}}
\newcommand{\Address}[1]{\begin{center}\small\sf #1 \end{center}}

\DeclareGraphicsExtensions{.eps,.jpg}

\renewenvironment{abstract}{\section*{Abstract}\normalsize\sf}{}
\newcommand{\References}[1]{\begin{flushleft}{\large References\\}\vspace*{2mm}\small #1 \end{flushleft}}

\newcommand{\chapterCoAst}[2]{\chapter[\sf\normalsize #1\\ \footnotesize \hspace*{5mm}by #2 \sf\normalsize][]{#1\\}\rhead[\fancyplain{}{\sf\footnotesize \center{#1}}]{\fancyplain{}{\sffamily\thepage}}\lhead[\fancyplain{\kopf}{\sffamily\thepage}]{\fancyplain{\kopf}{\sf\footnotesize \center{#2}}}}

\newcommand{\figureCoAst}[5]{\begin{figure}[#4]
\centering
\includegraphics*[#5]{#1}
\caption{#2}
\label{#3}
\end{figure}}

\renewcommand{\chaptername}{}